\documentclass[preprint,aps,showpacs,floatfix]{revtex4}
\usepackage{graphicx}
\usepackage{graphics}
\usepackage{psfrag}
\usepackage{color}      
\usepackage{subfigure}

\begin{document}

\title{The approach to vortex reconnection}
\author{Richard Tebbs$^1$}
\author{Anthony J. Youd$^1$}
\author{Carlo F. Barenghi$^1$}

\affiliation{$^1$ School of Mathematics and Statistics, 
Newcastle University, Newcastle upon Tyne, NE1 7RU, England, UK}

\date{\today}
\begin{abstract}
We present numerical solutions of the Gross--Pitaevskii equation
corresponding to reconnecting vortex lines. We determine the
separation of vortices as a function of time during the
approach to reconnection, and study the formation of pyramidal
vortex structures. Results are compared with analytical work
and numerical studies based on the vortex filament method.
\end{abstract}
\pacs{
67.25.dk, 
47.37.+q  
03.75.Kk  
}
\keywords{vortices,condensate,reconnection}
\maketitle

\newcommand{\vv}{\mathbf{v}}
\newcommand{\xhat}{\hat{\mathbf x}}
\newcommand{\rr}{\mathbf{r}}
\newcommand{\vvv}{\mathbf{\bar v}}
\newcommand{\etal}{{\it et al.}}

%

\section{Motivation}
\label{sec:1}

The discrete nature of vorticity in quantum fluids (superfluid $^4$He,
$^3$He and atomic Bose--Einstein condensates) makes it possible to
give physical meaning to vortex reconnections. 
The importance of vortex reconnections in turbulence cannot be
overestimated. Reconnections randomise the velocity field 
thus sustaining the turbulent vortex tangle in a steady state
\cite{Schwarz} and changing the flow's topology \cite{Poole}.
By triggering a Kelvin wave cascade
\cite{Svistunov-cascade,Kivotides-cascade,Vinen-cascade,Lvov-Nazarenko-arXiv,Kozik-Svistunov-JLTP-2009}, 
reconnections are responsible
for the decay of turbulent kinetic energy at very low temperatures
\cite{Vinen-Niemela}.
The dependence of the vortex reconnection frequency on
the observed vortex line density is therefore an important quantity
\cite{Lammi,Tsubota}.

In quantum fluids, the existence of vortex reconnections was 
first conjectured by Schwarz \cite{Schwarz}; it was then theoretically 
demonstrated by Koplik and Levine
\cite{Koplik}, and experimentally observed at the University of Maryland
\cite{Bewley,Paoletti}
using micron-size solid hydrogen tracers trapped
in superfluid vortex lines. More precisely, the Maryland
investigators inferred the
existence of reconnections by the sudden movement of trapped
particles. In doing so they relied on numerical calculations 
by de Waele and Aarts
\cite{Aarts},
who reported that the distance between two vortices, $\delta$, varies
as 
\begin{equation}
\delta(t) = (\kappa/2 \pi)^{1/2}\sqrt{t_0-t},
\label{eq:Aarts}
\end{equation}

\noindent
where $\kappa$ is the
quantum of circulation, $t$ is time and $t_0$ is the time of reconnection.
This $(t_0-t)^{1/2}$ scaling agrees with the approximate analytic
solution of the nonlinear Schroedinger equation 
(also called the Gross--Pitaevskii
equation, or GPE for short) found by Nazarenko and West \cite{West}.
De Waele and Aarts also claimed that, on their way toward a reconnection, 
the vortices
form a universal pyramidal cusp structure which is independent of the
initial condition. 

The numerical calculations of de Waele
and Aarts \cite{Aarts} were based on the vortex filament model 
pioneered by Schwarz \cite{Schwarz}, which
assumes that the vortex core is much smaller than any other
length scale in the problem. 
The vortex filament model thus breaks down in the vicinity
of a reconnection. It is therefore instructive to repeat the calculation
seeking full numerical solutions of the GPE, which does not suffer
such limitation and is a well-known model of superfluidity.
This is indeed what we plan to do in this paper, together with revisiting the
claim of universality of the reconnecting pyramidal cusp geometry.

\section{Model}
\label{sec:2}

The single-particle complex
wavefunction $\psi(\rr,t)$ for $N$ bosons of mass $m$
obeys the 3-dimensional time dependent GPE

\begin{equation}
i\hbar \frac{\partial \psi}{\partial t}=
-\frac{\hbar^2}{2m} \nabla^2 \psi - E \psi + V_0 \psi \vert \psi \vert^2,
\label{eq:dimGPE}
\end{equation}

\noindent
where ${\bf r}$ is the position, 
$V_0$ is the strength of the delta function interaction between
the bosons,  and $E$ is the single particle energy. The wavefunction is
normalised by the 
condition that $\int \vert \psi \vert^2 dV=N$. If we let 
$\tau=\hbar/(2E)$ be the
unit of time, $a=\hbar/\sqrt{2mE}$ the unit of length, and 
$\psi_0=\sqrt{E/V_0}$
the unit of $\psi$, we obtain the following dimensionless GPE 

\begin{equation}
-2 i \frac{\partial \psi}{\partial t}=
 \nabla^2 \psi + (1 - \vert \psi \vert^2)\psi.
\label{eq:GPE}
\end{equation}

\noindent
We solve Eq.~\ref{eq:GPE} in an $(x,y,z)$ periodic box, using
fourth order centred differences and Runge--Kutta--Fehlberg adaptive
time stepping. +he typical numerical resolution is $100 \times 300 \times 120$ 
for example for the first calculation described in Sec.~\ref{sec:3}. 
More details of the numerical technique and the method used to
set up initial conditions will be described elsewhere \cite{Helm}.

\section{\bf Ring--ring reconnections}
\label{sec:3}

\begin{figure}[!h]
  \centering
\includegraphics[scale=0.2]{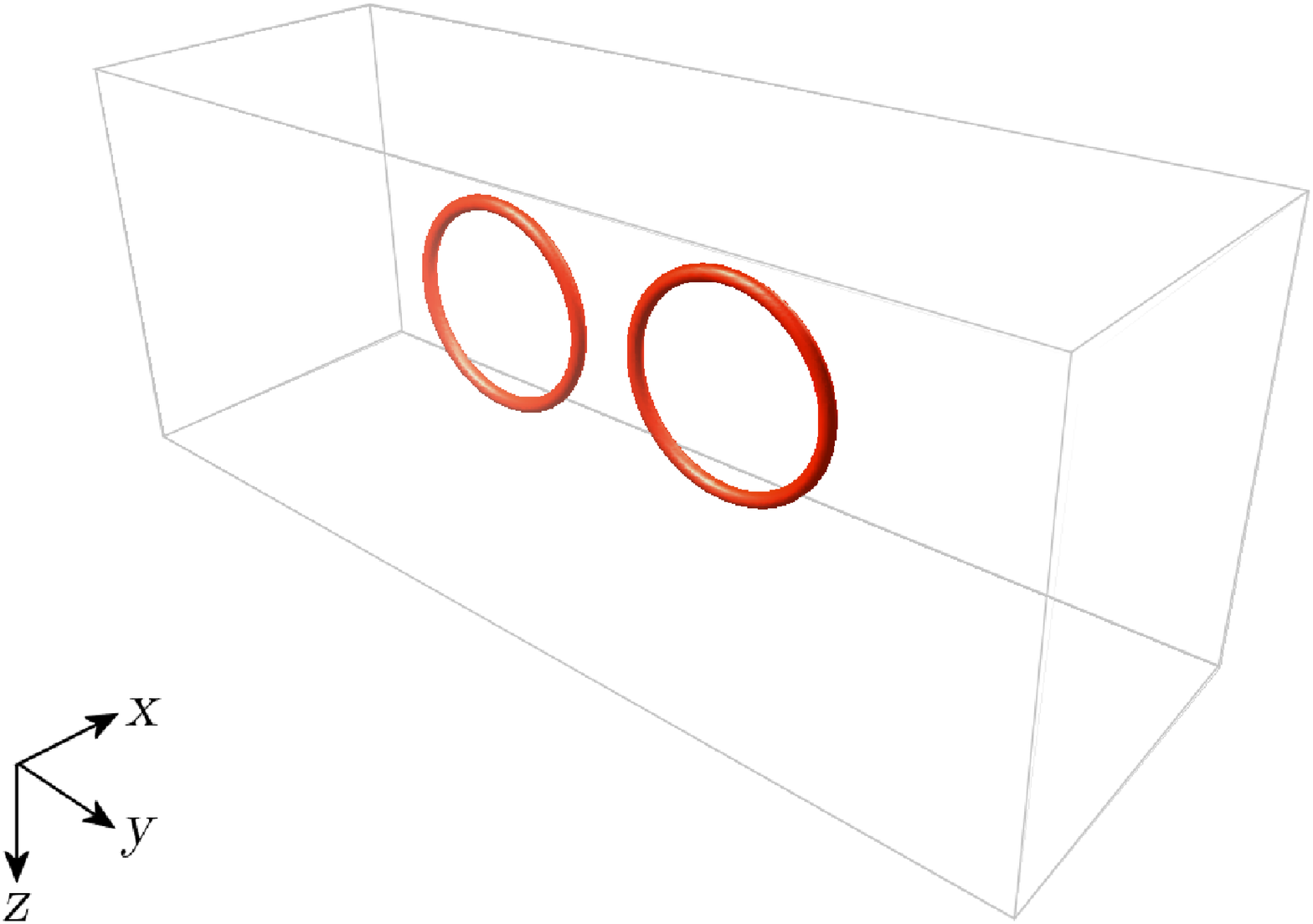}
\includegraphics[scale=0.2]{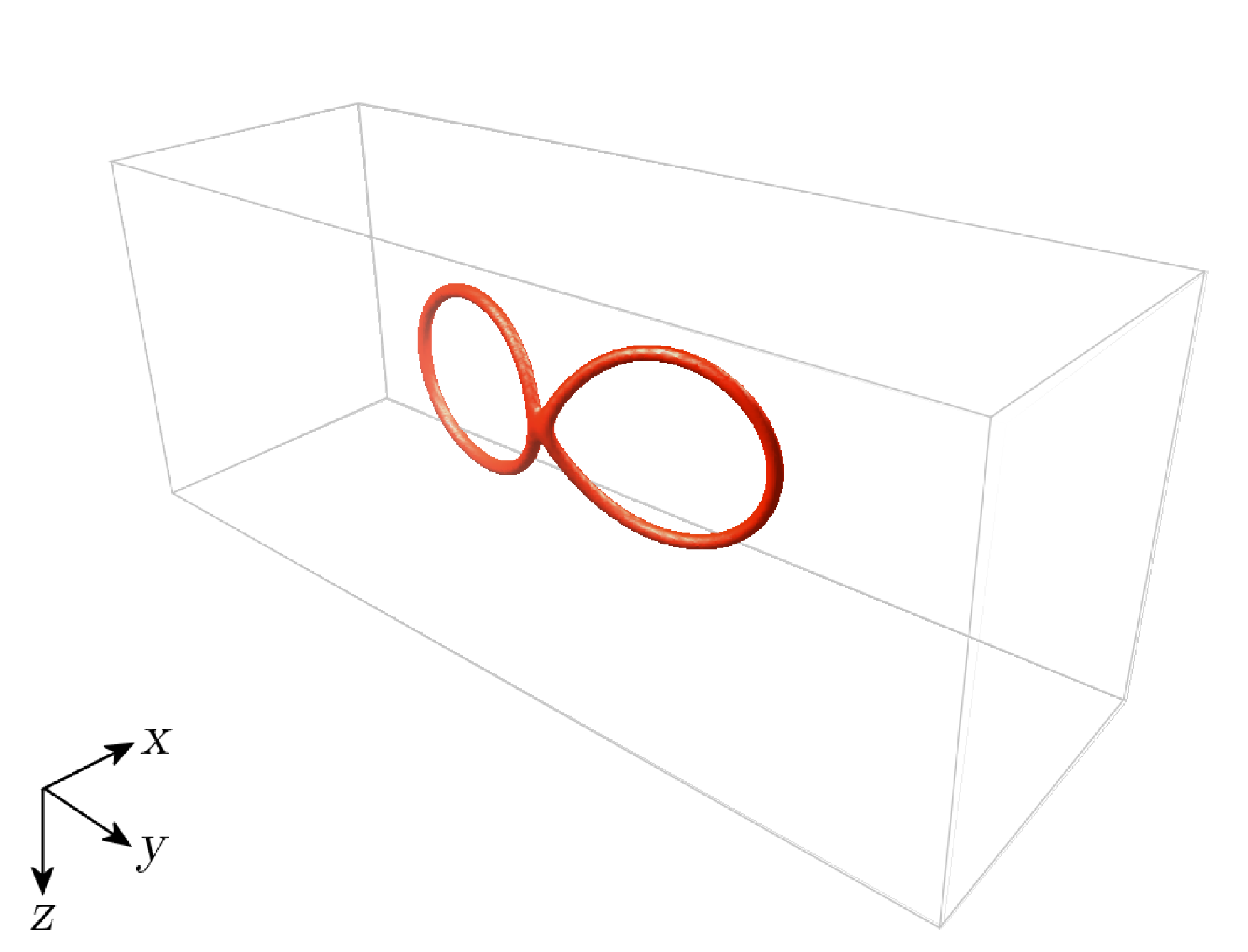}
\caption{\label{fig:1}(Colour online). Ring--ring reconnection.
Isosurfaces of $\vert \psi \vert^2=0.7$ (dimensionless units)
showing two rings moving along the positive $x$ direction
at (dimensionless) times
$t=0$ (left) and $t=81$ right. Note the formation of the
reconnecting cusp.}
\end{figure}

In a first set of numerical experiments, we consider the same planar
vortex rings 
configuration of de Waele and Aarts \cite{Aarts}. 
Two vortex rings of (dimensionless) radius $R=30$ are located on the
$y, z$ plane, initially separated by the (dimensionless)
distance $\delta(0)=20$, and moving in the positive $x$ direction.
Fig.~\ref{fig:1} shows that, during the evolution,
the rings turn into each other.
The pyramidal cusp which forms 
(and which eventually will lead to a vortex reconnection)
trails behind the rest of the moving vortex configuration.

\begin{figure}
  \centering
\includegraphics[angle=-90,width=0.7\textwidth]{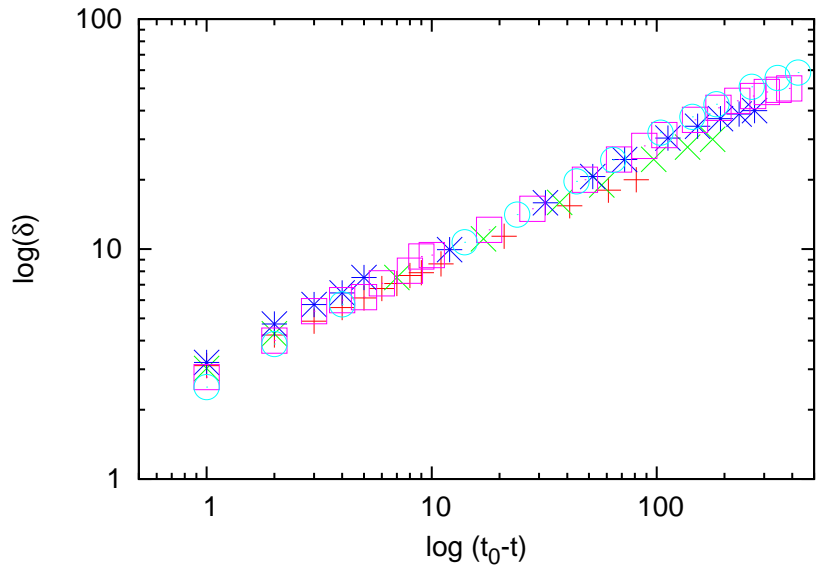}
\caption{\label{fig:2}
(Colour online). Reconnections of vortex rings. Log-log plot
of dimensionless distance $\delta(t)$ 
between the rings vs dimensionless time $t$, for various
initial distances $\delta(0)=20$ (red crosses), $30$ (green diagonal crosses),
$40$ (blue asterisks), $50$ (violet squares) and $60$ (pale blue circles).
}
\end{figure}

By symmetry, the reconnection is on the $z=0$ plane, which facilitates
the computation of the
minimum distance between the vortex rings, $\delta(t)$, as a function
of time $t$, until, at $t=t_0$, the rings reconnect. We repeat the
calculation for different initial separations $\delta(0)=30$,
$40$, $50$ and $60$. Fig.~\ref{fig:2} shows a log-log plot of
$\delta(t)$ vs $t$ and confirms that indeed
$\delta(t) \sim (t_0-t)^{1/2}$, in agreement with Eq.~\ref{eq:Aarts}.

We fit our $\delta(t)$ vs $t$ data to the form 

\begin{equation}
\delta(t)=A \sqrt{(t_0-t)}[1+c(t_0-t)],
\label{eq:fit}
\end{equation}

\noindent
where $A$ and $c$ are parameters.
We find that $A\gg c$, and that $A$ increases and $c$ decreases for increasing
$\delta(0)$. More precisely we obtain
$A=2.66$, $2.78$, $3.18$, $3.43$ and $3.63$, and
$c=2 \times 10^{-3}$, $1.06 \times 10^{-3}$, 
$0.87 \times 10^{-3}$, $0.67 \times 10^{-3}$ and
$0.52 \times 10^{-3}$, respectively for 
$\delta(0)=20$, $30$, $40$, $50$ and $60$. The errors are approximately
$\pm 0.08$ for  $A$ and $\pm 0.3 \times 10^{-3}$ for $c$, and arise 
mainly from the inaccuracy in localising the axes of the vortex lines.  

\begin{figure}
  \centering
\includegraphics[angle=0,width=0.5\textwidth]{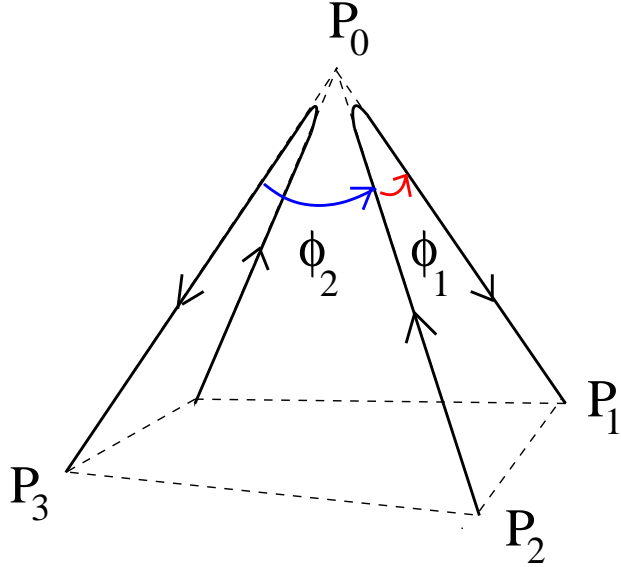}
\caption{\label{fig:3}
(Colour online). Pyramidal reconnection cusp and definition of the
angle $\phi_1$ (which a vortex makes with itself) and 
the angle $\phi_2$ (which a vortex makes with the other
vortex).
}
\end{figure}

\begin{figure}
  \centering
\includegraphics[angle=-90,width=0.5\textwidth]{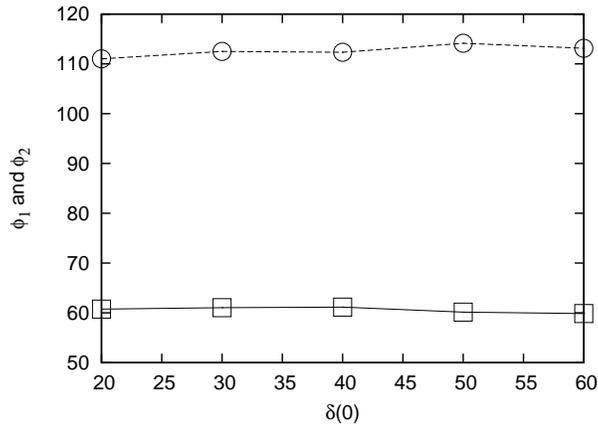}
\caption{\label{fig:4}
Reconnections of vortex rings. 
Reconnecting angles $\phi_1$ (top, circles) and
$\phi_2$ (bottom, squares) 
vs initial distance $\delta(0)$. The lines are to guide the eye.
}
\end{figure}

The Maryland group reported their results in the
(dimensional) form \cite{Paoletti} 

\begin{equation}
\delta_d(t_d)=A_M\sqrt{\kappa_d (t_{d0}-t_d)}[1+c_M(t_{d0}-t_d)],
\label{eq:UM}
\end{equation}

\noindent
(where we introduced the subscript ``d'' to stress 
that a quantity is dimensional),
where $A_M \approx 1.2$ and $c_M \approx 0.5~\rm s^{-1}$ 
are their fitting parameters. The numerical results of de Waele
and Aarts \cite{Aarts} correspond to $A_M=1/\sqrt{2 \pi}\approx 0.4$.

Since dimensional and dimensionless length, time and circulation are
related by $\delta=\delta_d/a$, $t=t_d/\tau$ and $\kappa=\kappa_d/(a^2/\tau)$
(where $\kappa_d=10^{-3}~\rm cm^2/s$ and $\kappa=2 \pi$ are respectively
the dimensional and dimensionless circulations),
our (dimensionless) results 
agree with Maryland's (dimensional) results 
if $A=A_M \sqrt{\tau \kappa_d/a^2}= A_M \sqrt{2 \pi} \approx 3$ 
and $c=\tau c_M\approx 10^{-13} \approx 0$,
which is indeed the case (the value of $E$ needed to get $\tau$ 
to compare $c$ with $c_M$ can be 
estimated from the definition of the healing length $a$ and the fact that
$a\approx 10^{-8}~\rm cm$). 

Following de Waele and Aarts \cite{Aarts}, we define the intravortex
angle $\phi_1$ and the intervortex angle $\phi_2$ of the pyramidal
cusp formed by the two vortices as they approach the reconnection.
These angles are computed from the dot products of the vector 
${\bf r}_2$ with the vectors ${\bf r}_1$ and ${\bf r}_3$, where 
${\bf r}_1={\bf P}_0-{\bf P}_1$, ${\bf r}_2={\bf P}_0-{\bf P}_2$, and
${\bf r}_3={\bf P}_0-{\bf P}_3$ as shown in Fig.~\ref{fig:3}. 
If the total angle $\phi_{tot}=2(\phi_1+\phi_2)$ is equal to $360^{\circ}$,
then the reconnection is ``flat.'' If $\phi_{tot}<360^{\circ}$ the vortices
form a pyramid.

Fig.~\ref{fig:4} shows that reconnecting angles are 
approximately constant, $\phi_1 \approx 112^{\circ}$ and
$\phi_2 \approx 61^{\circ}$,
independently of the initial distance between the vortex rings,
again in agreement with the finding of de Waele and Aarts \cite{Aarts},
although their numerical
values are slightly different:
their $\phi_1$ increases from $115^{\circ}$ to $135^{\circ}$ over a change of
two orders of magnitude of the initial distance. whereas
$\phi_2 \approx 25^{\circ}$.

\begin{figure}
  \centering
\includegraphics[angle=0,width=0.3\textwidth]{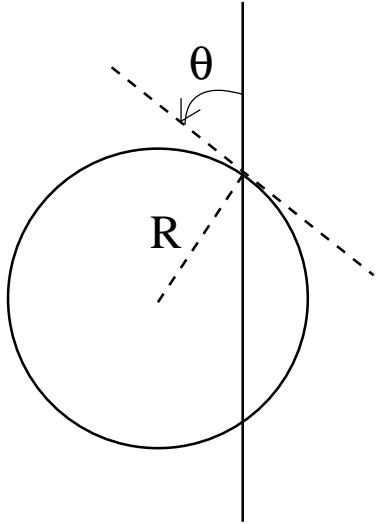}
\caption{\label{fig:5}
Initial vortex ring--line configuration
to define the angle $\theta$.
}
\end{figure}

\begin{figure}
  \centering
\includegraphics[angle=-90,width=0.7\textwidth]{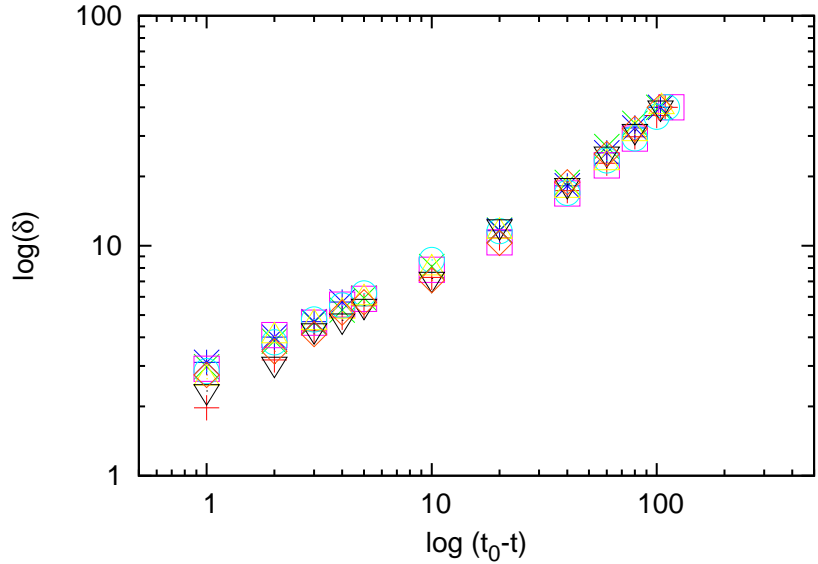}
\caption{\label{fig:6}
(Colour online). Ring--line reconnections. Log-log plot
of (dimensionless) distance $\delta(t)$ 
between the rings vs (dimensionless) time $t$, for
initial angles 
$\theta=60^{\circ}$ (red crosses), 
$75^{\circ}$ (green diagonal crosses),
$90^{\circ}$ (blue asterixes),
$105^{\circ}$ (violet squares),
$120^{\circ}$ (light blue circles),
$135^{\circ}$ (yellow pointing up triangles),
$150^{\circ}$ (brown pointing down triangles) and
$160^{\circ}$ (orange diamonds).
}
\end{figure}

\section{Ring--line reconnections}
\label{sec:4}

In a second set of numerical experiments, we consider the
interaction of a vortex line (set in the
middle of the computational box and aligned in the $z$
direction) with a vortex ring (located in the $y,z$ plane,
with centre at $y=D$) which moves in the positive $x$
direction). The geometry
of the initial configuration can be parametrised by the angle
$\theta$ between the vortex ring and the vortex line, see
Fig.~\ref{fig:5}; $\theta$ ranges from $\theta=0^{\circ}$
(the vortex ring passes to the left of the line) to $\theta=180^{\circ}$
(the vortex ring passes to the right of the line).

\begin{figure}
  \centering
\includegraphics[angle=-90,width=0.6\textwidth]{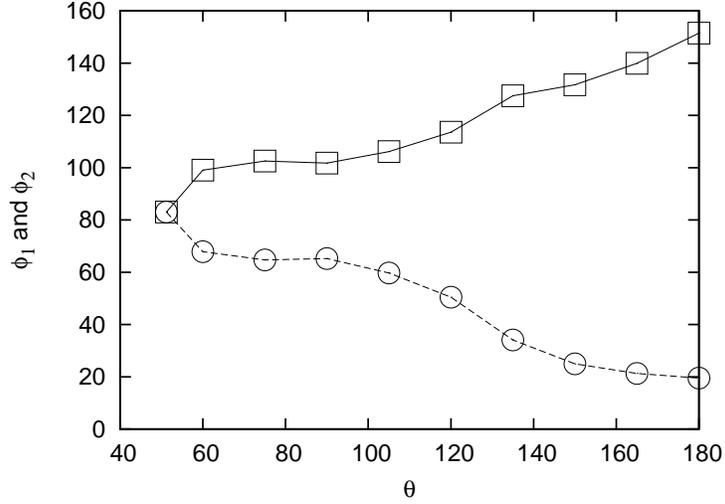}
\caption{\label{fig:7}
Ring--line reconnections. Reconnecting
angles $\phi_1$ (top, squares) and $\phi_2$ (bottom, circles)
vs initial angle $\theta$ (all angles in degrees). 
The lines are to guide the eye.
}
\end{figure}

\begin{figure}
  \centering
\includegraphics[angle=-90,width=0.6\textwidth]{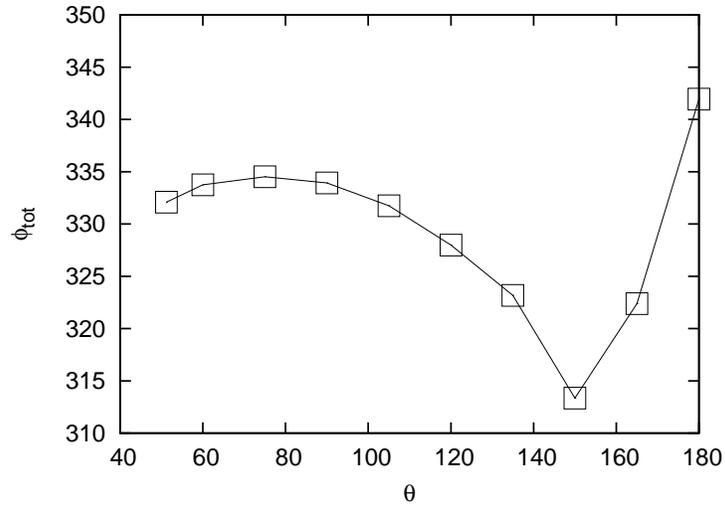}
\caption{\label{fig:8}
Ring--line reconnections. Total cusp angle
$\phi_{tot}$ vs
initial angle $\theta$ (all angles in degrees). 
The line is to guide the eye.
}
\end{figure}

Fig.~\ref{fig:6} shows that, for ring--line reconnections,
the distance between the vortices is very similar to what is
determined for ring--ring reconnections. Fitting as in Eq.~\ref{eq:fit}
we have $A=2.15$, $2.28$, $2.28$, $2.02$, $2.05$, $2.01$, $2.27$
and $2.25$, and 
$c=-7.2 \times 10^{-3}$,
$-7.2 \times 10^{-3}$,
$-6.9 \times 10^{-3}$,
$-7.2 \times 10^{-3}$,
$-7.6 \times 10^{-3}$,
$-8.9 \times 10^{-3}$,
$-7.0 \times 10^{-3}$ and
$-7.2 \times 10^{-3}$ respectively for initial angles
$\theta=60^{\circ}$, 
$75^{\circ}$,
$90^{\circ}$,
$105^{\circ}$,
$120^{\circ}$,
$135^{\circ}$,
$150^{\circ}$ and
$160^{\circ}$. 

In contrast to what happens in ring--ring
reconnections, there  is less evidence for a universal
pyramidal cusp. Fig~\ref{fig:7} shows that $\phi_1$
and $\phi_2$ depend on $\theta$, that is to say, on the
initial geometry. Fig~\ref{fig:8} shows that the total angle
$\phi_{tot}$ (which would be $360^{\circ}$ for a ``flat''
reconnection), is not constant.

\begin{figure}[ht]
    \label{plottop}
    \begin{center}
        \subfigure[t=0]{%
            \label{plottop-first}
            \includegraphics[width=0.39\textwidth]{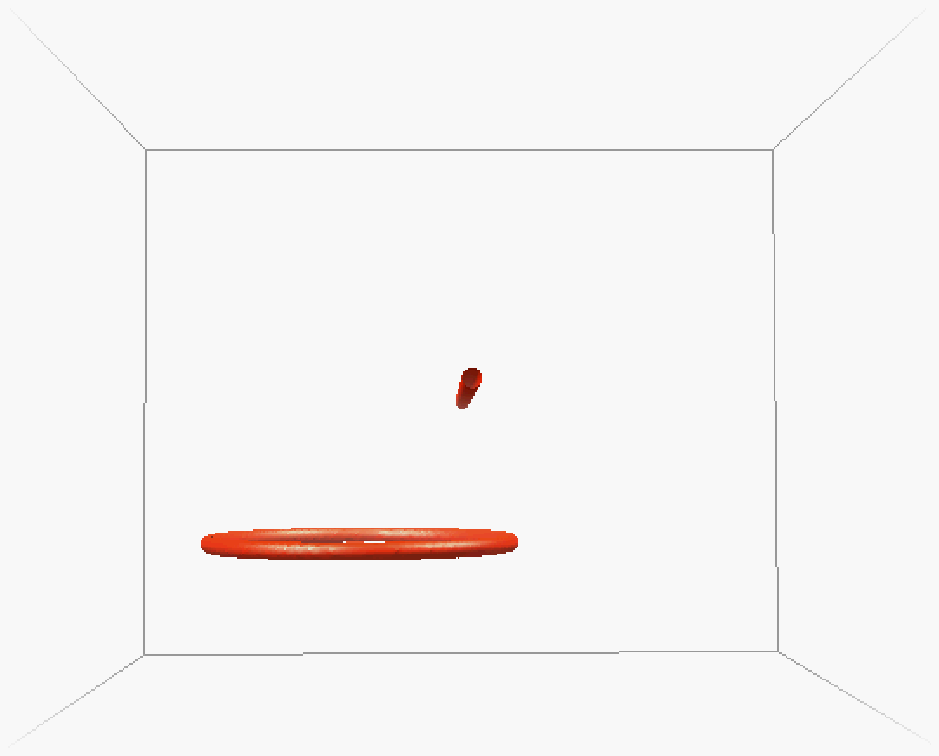}
        }
        \subfigure[t=22]{%
           \label{plottop-second}
            \includegraphics[width=0.39\textwidth]{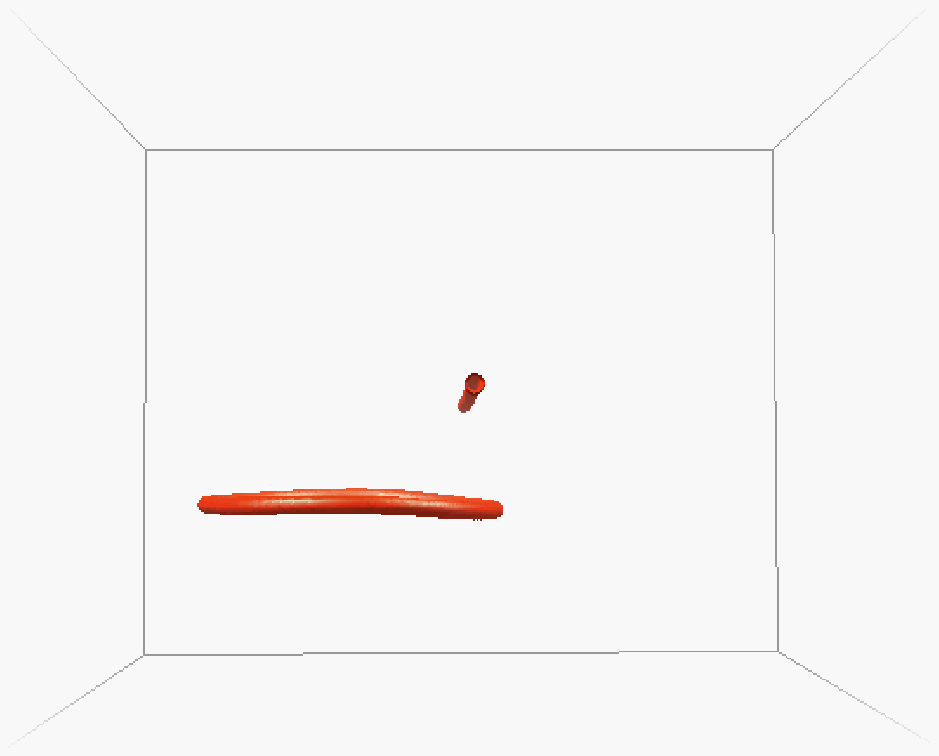}
        }
        \\
        \subfigure[t=44]{%
            \label{plottop-third}
            \includegraphics[width=0.39\textwidth]{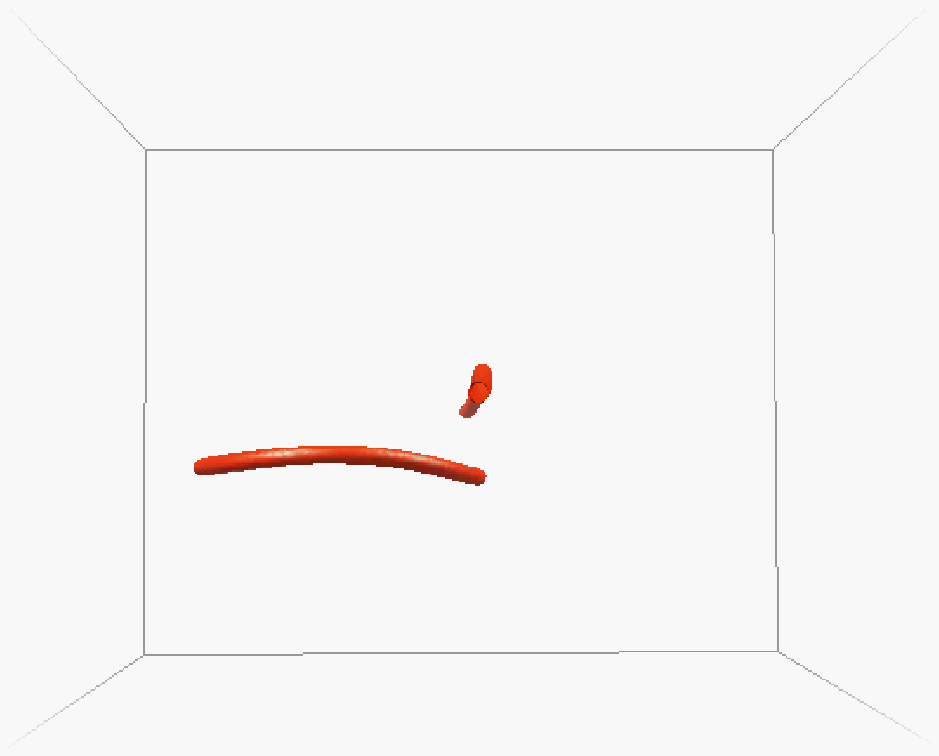}
        }
        \subfigure[t=66]{%
          \label{plottop-fourth}
            \includegraphics[width=0.39\textwidth]{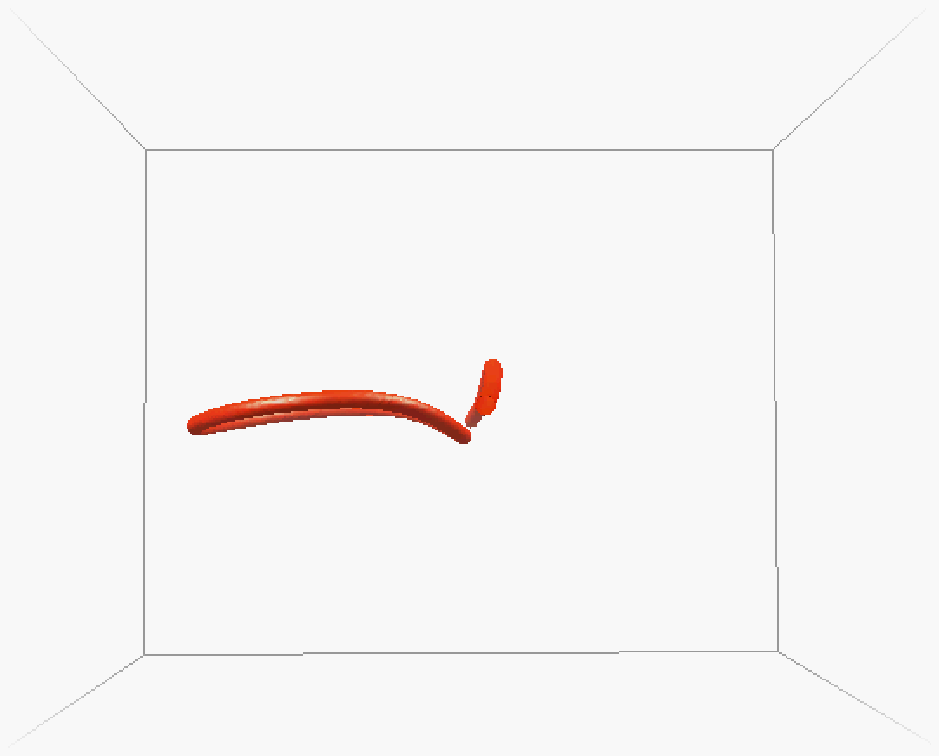}
        }
        \subfigure[t=88]{%
          \label{plottop-fourth}
            \includegraphics[width=0.39\textwidth]{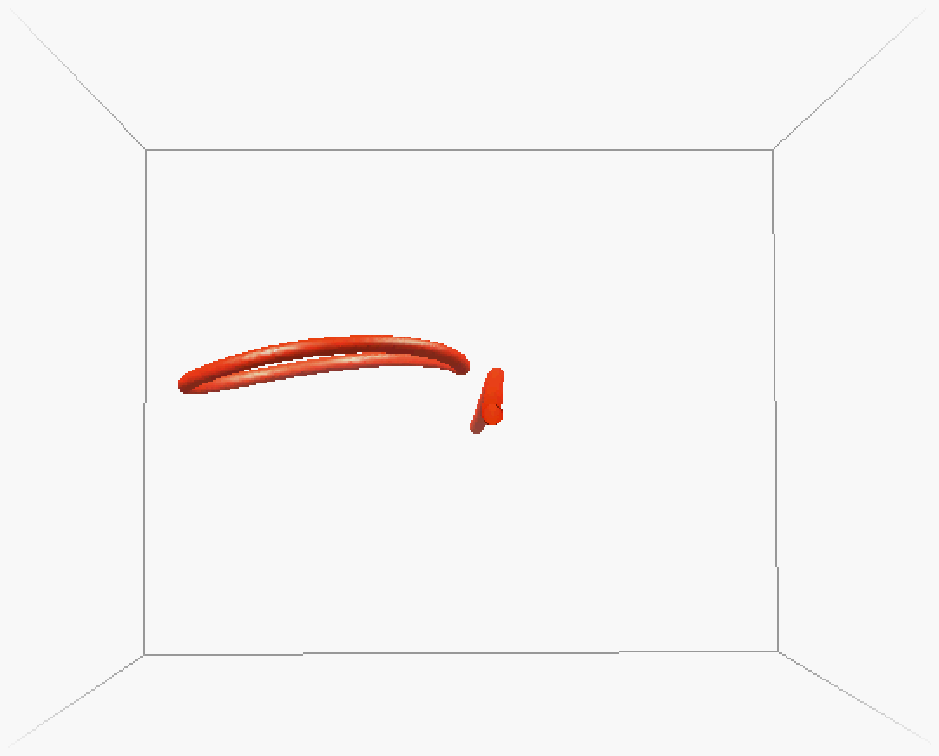}
        }
        \subfigure[t=110]{%
          \label{plottop-fourth}
            \includegraphics[width=0.39\textwidth]{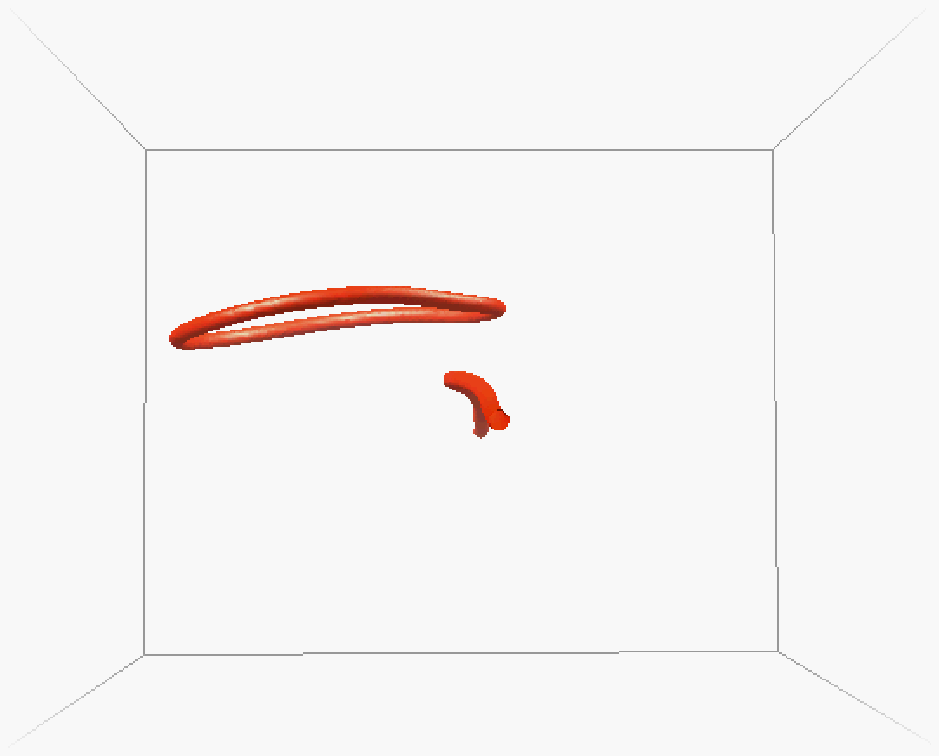}
        }
        \end{center}
    \caption{\label{fig:9}
(Colour online). Ring--line reconnection
for $\theta=45^{\circ}$, seen from the bottom, looking up the $z$ axis.
Isosurfaces of $\vert \psi \vert^2=0.7$ (dimensionless units) at
(domensionless) times $t=50$, $70$, $90$ and $110$.
The vortex ring (in the $y,z$ plane) moves in the positive $x$ direction
past the vortex line (aligned along $z$) without reconnecting.
}
\end{figure}

It is interesting to note that,
if $\theta<50^{\circ}$, when the vortex
ring is sufficiently close to the vortex line, the part of the vortex
ring which is near the vortex and the part of the vortex line 
which is near the ring are almost parallel; since they have
the same circulation, they tend to rotate about each other.
This motion perturbs and deflects the vortex ring, which passes
past the (perturbed) vortex line without any reconnection. 
Because of this curious effect,
shown in Fig.~\ref{fig:9},  we have no data for 
$\theta<50^{\circ}$ in Figs.~\ref{fig:7} and \ref{fig:8}.

\section{Conclusions}
\label{sec:5}

By numerically finding nonlinear solutions of the governing GPE,
we have confirmed that, as vortex filaments approach
a reconnection, their distance scales as $\delta \sim (t_0-t)^{1/2}$,
as predicted by Nazarenko and West \cite{West}. By fitting our
$\delta$ vs $t$ data to Eq.~\ref{eq:fit}, we find better
quantitative agreement with Maryland's observations \cite{Paoletti}
than obtained with the vortex filament method \cite{Aarts}.

Finally, we find that reconnections involve the formation of
a pyramidal cusp. However, on the length scales which we can explore
using the GPE (which are different from length scales described by the 
vortex filament method),  the angles of this cusp are not universal
as previously claimed \cite{Aarts}.


\end{document}